\documentclass[superscriptaddress, 12pt, aps, prl,  preprint, longbibliography, nofootinbib]{revtex4-1}
\usepackage{graphicx}
\usepackage{epstopdf}
\usepackage{color}
\usepackage{amsmath, amssymb}

\begin{document}
\title{Comment on "Mineral–water reactions in Earth's mantle: Predictions from Born theory and ab initio molecular dynamics" by Fowler \textit{et al.} 2024 (Geochim. Cosmochim. Acta 372, 111-123)}
\author{Jiajia Huang}
\affiliation{Department of Physics, The Hong Kong University of Science and Technology, Clear Water Bay, Hong Kong, P. R. China.}
\author{Ding Pan}
\email{dingpan@ust.hk}
\affiliation{Department of Physics, The Hong Kong University of Science and Technology, Clear Water Bay, Hong Kong, P. R. China.}
\affiliation{Department of Chemistry, The Hong Kong University of Science and Technology, Clear Water Bay, Hong Kong, P. R. China}

\date{\today}

\begin{abstract}
    This comment addresses discrepancies in dielectric constant ($\varepsilon_0$) calculations of water under extreme conditions ($\sim$ 10 GPa and 1000 K) between Fowler et al.'s recent study [Geochim. Cosmochim. Acta 372, 111-123 (2024)] and the earlier work by Pan et al. [Proc. Natl. Acad. Sci. 110, 6646–6650 (2013)]. Through reproduced ab initio molecular dynamics (AIMD) simulations using the CP2K code with extended duration and identical system size, we validate that Pan et al.'s original results (39.4) are well-converged, contrasting with Fowler et al.'s reported value of 51. The observed discrepancy cannot be attributed to simulation duration limitations, but rather to methodological differences in dipole moment calculation. Our analysis highlights critical issues in the treatment of dipole moment fluctuations in periodic systems within the framework of modern theory of polarization. This clarification has significant implications for modeling mineral-water interactions in Earth's mantle using Born theory.
\end{abstract}
\maketitle

In Ref. \cite{Fowler2024Mineral-water}, Fowler \textit{et al.} applied ab initio molecular dynamics (AIMD) simulations to calculate the dielectric constant of water, $\varepsilon_0$, up to 30 GPa and 3000 K. They used this key parameter in the Born model to study mineral-water interactions in the Earth's mantle.
They compared their results with a similar previous study by Pan \textit{et al.} \cite{Pan2013Dielectric}, and found good agreement at $\sim$ 1 and $\sim$ 6 GPa and 1000 K, but poor agreement at $\sim$ 10 GPa and 1000 K. The dielectric constant of water at $\sim$ 10 GPa and 1000 K is 39.4$\pm$0.5 in Pan \textit{et al.}'s original work \cite{Pan2013Dielectric}, while in Fowler \textit{et al.}'s work, $\varepsilon_0$ is 51 \cite{Fowler2024Mineral-water}.  They attributed this discrepancy to the short simulation time in Ref. \cite{Pan2013Dielectric}. 
However, in fact, Pan \textit{et al.} have further developed a neural network dipole model to calculate $\varepsilon_0$ more efficiently and achieved precision comparable to ab initio methods \cite{Hou2020Dielectric}. By integrating this model with machine learning force fields, they conducted molecular dynamics simulations with significantly extended timescales compared to Ref. \cite{Fowler2024Mineral-water}. 
At $\sim$ 10 GPa and 1000 K, the AIMD simulation time in Pan \textit{et al.}'s original work was about 25 ps \cite{Pan2013Dielectric}. In the work of Fowler \textit{et al.}, the simulation time was 150$\sim$400 ps  \cite{Fowler2024Mineral-water}. 
In the machine learning study, the simulation time exceeded 2,100 ps \cite{Hou2020Dielectric}, much longer than the simulation time in Fowler \textit{et al.}'s work \cite{Fowler2024Mineral-water}. 
Using this machine learning approach, the calculated dielectric constant of water at $\sim$ 10 GPa and 1000 K is 40.7$\pm$0.12 \cite{Hou2020Dielectric}, which is in excellent agreement with the value, 39.4$\pm$0.5, obtained from the original AIMD simulations \cite{Pan2013Dielectric}, indicating that $\varepsilon_0$ reported in Ref. \cite{Pan2013Dielectric} is well converged.
What's more, the simulation box used in the machine learning study contains 256 water molecules, surpassing the 128 molecules in Pan \textit{et al.} \cite{Pan2013Dielectric} and the 110 molecules in Fowler \textit{et al.} \cite{Fowler2024Mineral-water}.
Therefore, the discrepancy of $\varepsilon_0$ between Refs. \cite{Fowler2024Mineral-water} and \cite{Pan2013Dielectric} cannot be explained by the unconverged AIMD simulations.

We found that Fowler \textit{et al.} did not calculate the dipole moment of the simulation boxes properly. This oversight may lead to incorrect dipole moment fluctuations, potentially explaining why their reported dielectric constant of water is significantly different from Pan et al.'s original work at $\sim$ 10 GPa and 1000 K. To clarify this procedure, we provide a concise derivation for calculating $\varepsilon_0$ here. 
Both Refs. \cite{Fowler2024Mineral-water} and \cite{Pan2013Dielectric} applied Neumann's linear response theory to compute $\varepsilon_0$ using the fluctuations of the total dipole moment, $\Vec{M}$, except that Fowler \textit{et al.} ignored electronic polarizability \cite{Neumann1983Dipole, Neumann1984Computer}. The dielectric tensor, $\epsilon$, in the Gaussian unit is defined as
\begin{equation}\label{Eq:dielec_def}
    \epsilon=\frac{\vec{E}+4 \pi \vec{P}}{\vec{E}}=I+4 \pi \frac{\vec{M}}{V \vec{E}},
\end{equation}
where $\vec{E}$ is the macroscopic electric field, $\vec{P}$ is the polarization, 
$V$ is the volume of the simulation box with periodic boundary conditions.  
In the AIMD simulation with a canonical (i.e., NVT) ensemble, we aim to obtain the ensemble average:
\begin{equation}\label{Eq:ensemble_E}
    \langle\vec{M}\rangle_{\vec{E}}=\frac{\int \vec{M} e^{-\beta(H-\vec{M} \cdot \vec{E_0})} d p^N d q^N}{\int e^{-\beta(H-\vec{M} \cdot \vec{E_0})} d p^N d q^N},
\end{equation}
where $\beta$ is the thermodynamic beta, $H$ is the Hamiltonian of the system, $\vec{E_0}$ is the applied electric field, N is the number of atoms and the integral $\int...d p^N d q^N$ is over the whole phase space.
If the electric field is very weak, we can approximate $e^{\beta \vec{M} \cdot \vec{E}_0} \approx 1 + \beta \vec{M} \cdot \vec{E}_0 $ in the linear region, and Eq. \ref{Eq:ensemble_E} becomes
\begin{equation}\label{Eq:ensemble_E_linear}
    \langle\vec{M}\rangle_{\vec{E}}=\frac{\int \vec{M} e^{-\beta H}(1+\beta \vec{M} \cdot \vec{E_0}) d p^N d q^N}{\int e^{-\beta H}(1+\beta \vec{M} \cdot \vec{E_0}) d p^N d q^N}.
\end{equation}
The ensemble average of $\vec{M}$ at $\vec{E_0}=0$ gives
\begin{equation}\label{Eq:ensemble_0}
    \langle\vec{M}\rangle=\frac{\int \vec{M} e^{-\beta H} d p^N d q^N}{\int e^{-\beta H} d p^N d q^N}.
\end{equation}
We divide the numerator and denominator of Eq. \ref{Eq:ensemble_E_linear} by the partition function $\int e^{-\beta H} d p^N d q^N$, and substitute Eq. \ref{Eq:ensemble_0} into it:
\begin{align}
    \langle\vec{M}\rangle_{\vec{E}} &= \frac{\langle\vec{M}\rangle + \beta\langle\vec{M}(\vec{M} \cdot \vec{E_0})\rangle}{1+\beta\langle\vec{M}\rangle \cdot \vec{E_0}} \\
    &= (\langle\vec{M}\rangle+\beta\langle\vec{M}(\vec{M} \cdot \vec{E_0})\rangle) \cdot(1-\beta\langle\vec{M}\rangle \cdot \vec{E_0} ),
\end{align}
where we use again the linear approximation in the second step: $\frac{1}{1+\beta\langle\vec{M}\rangle \cdot \vec{E_0}} \approx 1-\beta\langle\vec{M}\rangle \cdot \vec{E_0}$.
We keep only the first order term of $\vec{E_0}$:
\begin{equation}\label{Eq:linear_M_E}
    \langle\vec{M}\rangle_{\vec{E}} = \beta\langle\vec{M} \vec{M}\rangle \cdot \vec{E_0}-\beta\langle\vec{M}\rangle\langle\vec{M}\rangle
    \cdot \vec{E_0},
\end{equation}
where $\vec{M}\vec{M}$ and $\langle\vec{M}\rangle\langle\vec{M}\rangle$ are two second rank tensors. 

In density functional theory, calculations with periodic boundary conditions as reported in Refs. \cite{Fowler2024Mineral-water, Pan2013Dielectric}, electrostatic interactions such as Hartree energies and Ewald sums are calculated in the reciprocal space. The Fourier component of the internal electric field at $\vec{G}=0$ is set to 0, a condition known as conducting boundary conditions according to Ref. \cite{Neumann1983Dipole}. Thus, $\vec{E_0}$ is equal to $\vec{E}$. 
After substituting $\vec{M}$ in Eq. \ref{Eq:dielec_def} by $\langle\vec{M}\rangle_{\vec{E}}$ in Eq. \ref{Eq:linear_M_E} , we obtain the dielectric tensor
\begin{equation}
    \epsilon = I + \frac{4 \pi \beta}{V} (\langle \vec{M}\vec{M}\rangle-\langle\vec{M}\rangle\langle\vec{M}\rangle ).
\end{equation}
For isotropic systems, the static dielectric constant is
\begin{equation}\label{Eq:epsilon_rigid}
    \varepsilon_0=\frac{1}{3} \operatorname{Tr}(\epsilon) = 1 + \frac{4 \pi \beta}{3V} (\langle \vec{M}^2\rangle-\langle\vec{M}\rangle^2),
\end{equation}
which is the equation used in Fowler \textit{et al.}'s work \cite{Fowler2024Mineral-water}.

In AIMD simulations, we employed the Born–Oppenheimer approximation, ensuring the electronic structure is always converged to the ground state at the temperature of $T = 0$ K, without any thermal fluctuations. Therefore, we need to consider the fluctuations of the electronic dipole moment $\vec{M_e}$ separately:
\begin{equation}\label{Eq:M_e-M_ion}
  \langle \vec{M}^2\rangle = \langle \vec{M}_{ion}^2\rangle + \langle \vec{M}_e^2\rangle,
\end{equation}
where the ionic dipole moment $\vec{M}_{ion}$ is uncorrelated with $\vec{M_e}$.
After substituting Eq. \ref{Eq:M_e-M_ion} into Eq. \ref{Eq:epsilon_rigid}, we finally have
\begin{equation}\label{Eq:epsilon_polarizable}
    \varepsilon_0=\frac{1}{3} \operatorname{Tr}(\epsilon) = \varepsilon_\infty + \frac{4 \pi \beta}{3V} (\langle \vec{M}_{ion}^2\rangle-\langle\vec{M}_{ion}\rangle^2),
\end{equation}
where $\varepsilon_\infty$ is the electronic (optical) dielectric constant \cite{Neumann1984Computer}. 
Fowler \textit{et al.} ignored $\varepsilon_\infty \sim 2.41$ at $\sim$10 GPa and 1000 K and used $\varepsilon_\infty=1$ in their calculations. While this choice partly contributes to their higher $\varepsilon_0$, it cannot fully explain the discrepancy. 

Note that Fowler \textit{et al.} used the Berry phase method to calculate the total dipole moment of the simulation box, $\vec{M}$ \cite{Fowler2024Mineral-water}, whereas Pan \textit{et al.} used the maximally localized Wannier functions (MLWF) \cite{Pan2013Dielectric}.
In principle, these two methods should give the consistent result if the simulation box is large enough \cite{Stengel2006Accurate}. 
However, According to the modern theory of polarization,  the polarization of a simulation box with periodic boundary conditions does not have a single value; instead, it takes on multiple values, which differ by the polarization quantum, i.e.,  the polarization change resulting from moving one charged particle by one unit cell with periodic boundary conditions \cite{Spaldin2012beginners}. This variability makes it difficult to determine a unique result from the equations discussed above, except when the system contains purely dipolar particles and the polarization or total dipole moment is uniquely defined.
Thus, Pan \textit{et al.} first assigned two hydrogen atoms and four MLWF centers to each oxygen atom to calculate the molecular dipole moment, and then summed over all the water molecules \cite{Pan2013Dielectric, li2025ab}.
Unfortunately, this procedure was missing in Fowler \textit{et al.}'s work \cite{Fowler2024Mineral-water}. 
Fowler \textit{et al.} used the CP2K code \cite{kuhne2020cp2k}, and in the code manual regarding "MOMENTS" clearly states that `` Note that the result in the periodic case might be defined modulo a certain period, determined by the lattice vectors. During MD, this can lead to \emph{jumps}" 
\footnote{The definition of moments in {CP2K}, \url{https://manual.cp2k.org/trunk/CP2K_INPUT/FORCE_EVAL/DFT/PRINT/MOMENTS.html}, Accessed: 2025-07-16}

\begin{figure}[ht]
    \centering
    \includegraphics[width=0.5\textwidth]{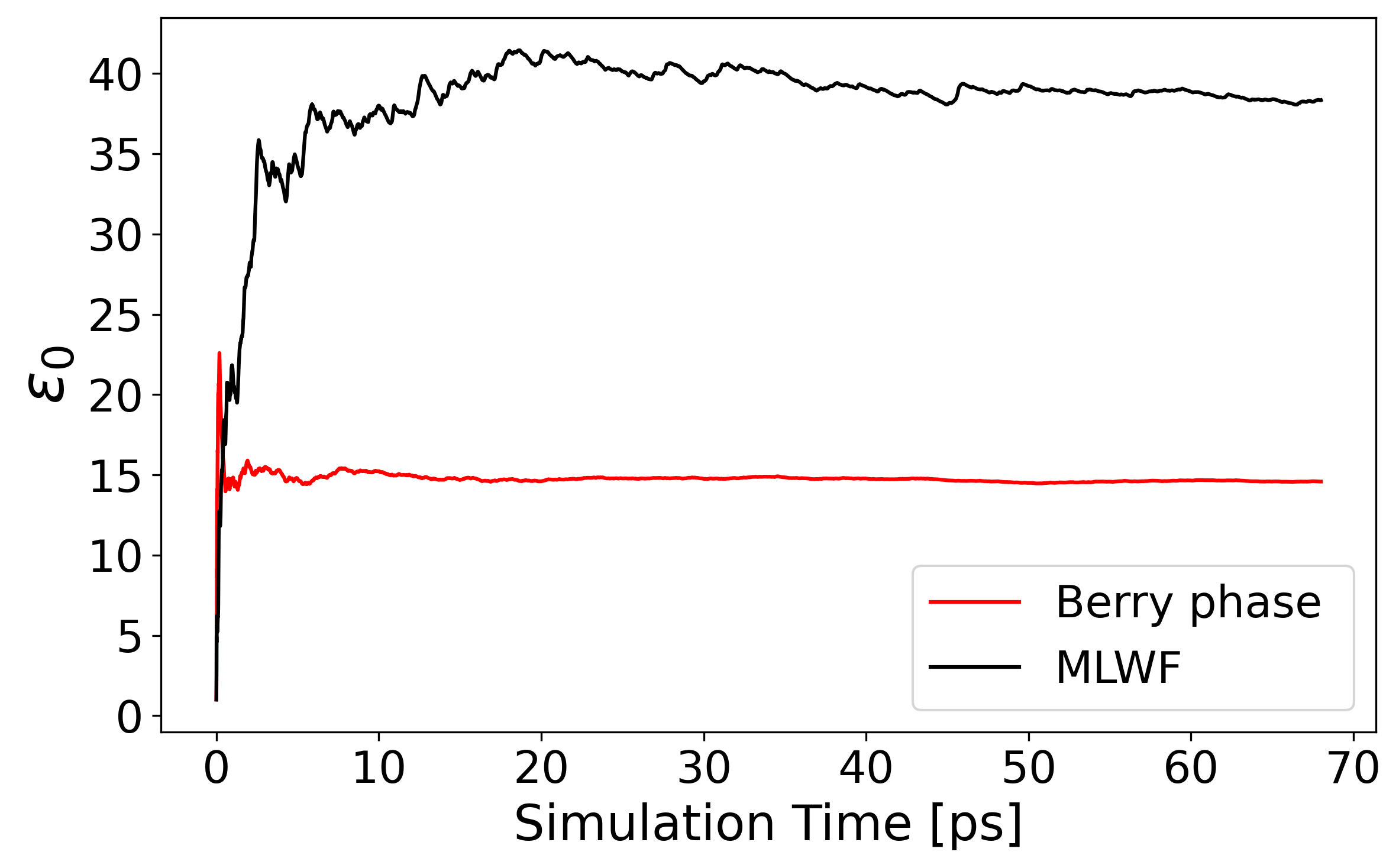}
    \caption{The static dielectric constant of water, $\epsilon_0$, as a function of simulation time,obtained by Berry phase (red) and MLWF (black) methods implemented in the CP2K code. In the MLWF method, we assigned two hydrogen atoms and four MLWF centers to each water molecule.}
    \label{fig:epsilon-wannier}
\end{figure}

\begin{figure}[ht]·
    \centering
    \includegraphics[width=0.5\textwidth]{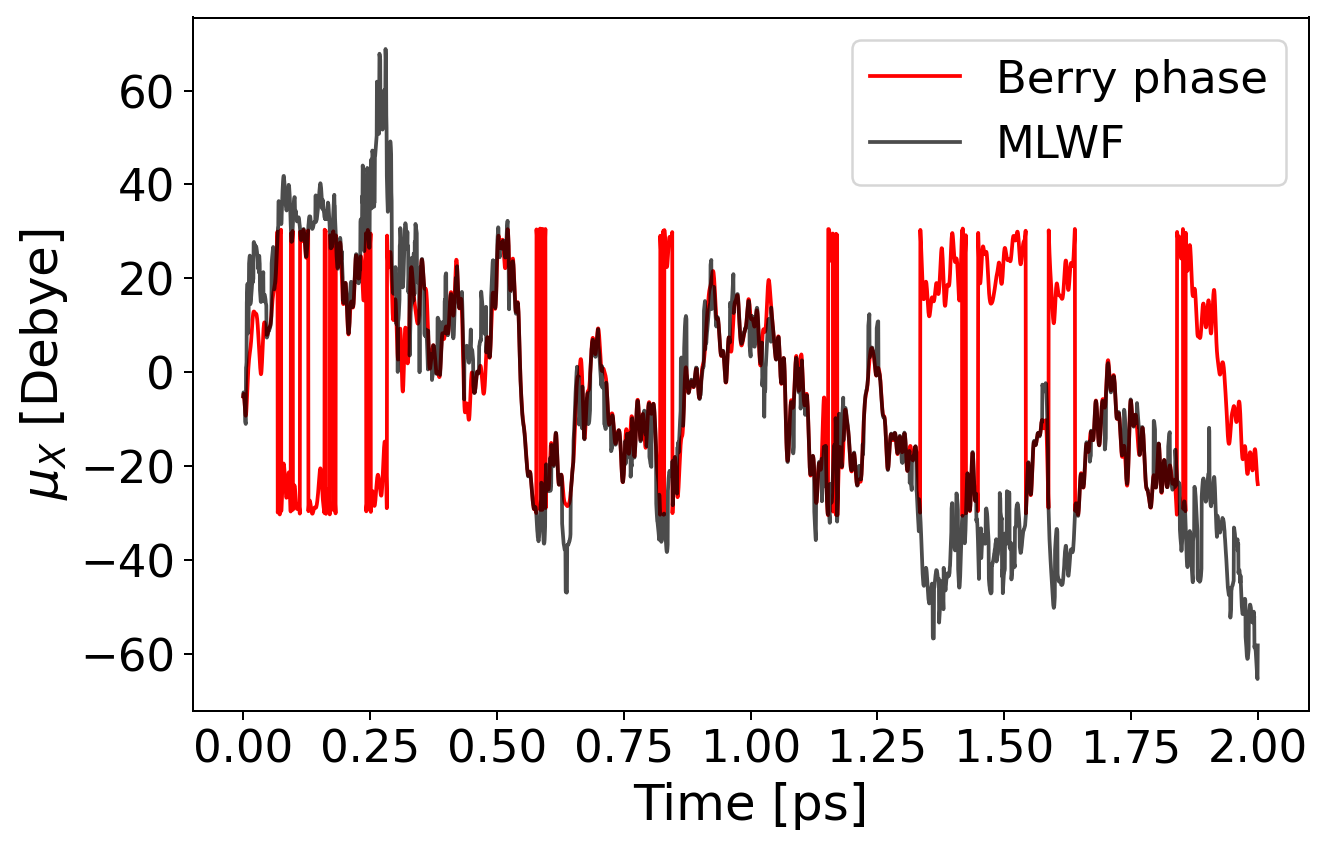}
    \caption{The dipole moment of the simulation box along the x-direction as a function of simulation time, calculated by the Berry phase (red) and MLWF (black) methods. The Berry phase method shows unphysical discontinuities.}
    \label{fig:mu_x}
\end{figure}

To verify our argument, we further performed the AIMD simulations at  $\sim$10 GPa and 1000 K using the CP2K code with input parameters identical to those reported by Fowler \textit{et al.} \cite{lord2024input}. Fig. \ref{fig:epsilon-wannier} shows that the dielectric constants calculated by the Berry phase method (14.6) and MLWF approach (38.3)  have significant discrepancies.
The Berry phase result failed to reproduce Fowler \textit{et al.}'s value of 51, while the MLWF result aligned with Pan \textit{et al.}'s original work. Both methods converged within 25 ps, confirming that the simulation time in Pan \textit{et al.}'s study was sufficient.
Further analysis of the x-direction dipole moment $\mu_x$ in Fig.\ref{fig:mu_x} shows unphysical discontinuities in the result obtained by the Berry phase method. The range of $\mu_x$ is limited between $\sim$$\pm$30 Debye, corresponding to the length of the simulation box in x direction multiplied by the elementary charge, $\sim$60 Debye. This confirms that the CP2K code artificially wraps the dipole moment within a single polarization quantum under periodic boundary conditions. In contrast, the MLWF approach in Pan \textit{et al.}'s original work can uniquely define the correct dipole moment for dipolar liquids.

Furthermore, Fowler \textit{et al.} ``observe that
mean square displacement (MSD) diverges for O and H (Fig. S4) with
increasing P, T. The implication is that H are increasingly delocalized (i.e. not bound to a particular O)" \cite{Fowler2024Mineral-water}. This means that there are ``delocalized" H$^+$ or H$_3$O$^+$ ions, implying that the total Hamiltonian can not be written as $H-\vec{M} \cdot \vec{E_0}$ in Eq. \ref{Eq:ensemble_E}, so eventually Eqs. \ref{Eq:epsilon_rigid} and \ref{Eq:epsilon_polarizable} do not work.
In these cases, the correct way to calculate the dielectric constant of liquids with such long-lived ions is documented in the literature (see, e.g., Refs. \cite{Caillol1986Theoretical, Sala2010Effects}).

In summary, Fowler \textit{et al.} reported the dielectric constant of water up to 30 GPa and 3000 K. However, they did not calculate the dipole moment of the simulation boxes properly, leading to incorrect dipole moment fluctuations even when there are no long-lived ions, so their obtained dielectric constant of water is incorrect at $\sim$ 10 GPa and 1000 K. 
While at higher pressures and temperatures, where water dissociates and long-lived ions are present, Eqs. \ref{Eq:epsilon_rigid} and \ref{Eq:epsilon_polarizable} do not work, calling into question the results derived from these equations in their work. The dielectric constant of water is a key parameter in the Born model for studying mineral-water interactions in Earth’s mantle. Inaccurate dielectric constants can result in erroneous ion solvation and mineral dissolution. 

\section{Declaration of competing interest}
The authors declare that they have no known competing financial
interests or personal relationships that could have appeared to influence
the work reported in this paper.

\section{Acknowledgments}
We thank Dimitri A. Sverjensky for reading our manuscript and for many valuable discussions.
This work was supported by the Research Grants Council of Hong Kong (Projects GRF-16301723 and GRF-16306621), and the National Natural Science Foundation of China /Research Grants Council of Hong Kong Joint Research Scheme (N\_HKUST664/24)

\bibliography{refs.bib}

\end{document}